%% file: main.tex
\documentclass[10pt,twocolumn,letterpaper]{article}

\usepackage{cvpr}
\usepackage{times}
\usepackage{epsfig}
\usepackage{graphicx}
\usepackage{amsmath}
\usepackage{amssymb}
\usepackage{longfbox}
\usepackage{hyperref}
\usepackage{apacite}

\usepackage{todo}
\usepackage{color}

\usepackage[caption=false,font=scriptsize]{subfig}
\usepackage[linesnumbered,lined,boxed,commentsnumbered]{algorithm2e}

\hypersetup{
    colorlinks = true,
    linkbordercolor = white,
    citecolor = blue
}

\cvprfinalcopy 

\def\cvprPaperID{7} 

\SetAlCapSkip{0.2cm}
\begin{document}

\title{Visual pathways from the perspective of cost functions and multi-task deep neural networks}

\author{H.Steven Scholte$^{1,2,\ast}$~~Max M. Losch$^{1,2,3,\ast}$~~Kandan Ramakrishnan$^3$\\Edward H.F. de Haan$^{1,2}$~~Sander M. Bohte$^4$\\
$^\ast$ Shared first author\\
$^1$Department of Psychology, University of Amsterdam, The Netherlands\\
$^2$Amsterdam Brain and Cognition, University of Amsterdam, The Netherlands\\
$^3$Informatics Institute, University of Amsterdam, The Netherlands\\
$^4$Machine Learning Group, CWI, Amsterdam, The Netherlands\\
{\tt\small \{h.s.scholte, m.m.losch, k.ramakrishnan, e.h.f.dehaan\}@uva.nl, s.m.bohte@cwi.nl}
}

\maketitle

\begin{abstract}
Vision research has been shaped by the seminal insight that we can understand the higher-tier visual cortex from the perspective of multiple functional pathways with different goals. In this paper, we try to give a computational account of the functional organization of this system by reasoning from the perspective of multi-task deep neural networks. Machine learning has shown that tasks become easier to solve when they are decomposed into subtasks with their own cost function. We hypothesize that the visual system optimizes multiple cost functions of unrelated tasks and this causes the emergence of a ventral pathway dedicated to vision for perception, and a dorsal pathway dedicated to vision for action. 
To evaluate the functional organization in multi-task deep neural networks, we propose a method that measures the contribution of a unit towards each task, applying it to two networks that have been trained on either two related or two unrelated tasks, using an identical stimulus set. Results show that the network trained on the unrelated tasks shows a decreasing degree of feature representation sharing towards higher-tier layers while the network trained on related tasks uniformly shows high degree of sharing. 
We conjecture that the method we propose can be used to analyze the anatomical and functional organization of the visual system and beyond. We predict that the degree to which tasks are related is a good descriptor of the degree to which they share downstream cortical-units.

\end{abstract}

\input{sections/section_1}
\input{sections/section_2}
\input{sections/section_3}
\input{sections/section_4}
\input{sections/section_5}

\section*{Acknowledgements}
MML is supported by a grant from the ABC, KR is supported by a grant from COMMIT and EHFdH by an ERC (339374 - FAB4V).

{\small
\bibliographystyle{apacite}
\bibliography{references}
}

\input{sections/appendix}

\end{document}

%% file: sections/section_1.tex
\section{Introduction}
The visual system is described as consisting of two parallel pathways. Research by Gross, Mishkin and colleagues, integrating insights from lesion \cite{Newcombe1969-yj} and anatomical studies \cite{Schneider1969-yc}, showed that these pathways emerge beyond the striate cortex with one involved in the identification of objects projecting ventrally, and the other involved in localization of objects, projecting to the parietal cortex \cite{Gross1977-xv,Mishkin1983-qi}. From the start of the dual-pathway theory, multiple pathways were believed to be computationally efficient \cite{Gross1977-xv}. Support for this idea comes from research using artificial networks with one hidden layer, showing that location and identity are better learned when units in the hidden layers are uniquely assigned to one of these functions \cite{Rueckl1989-ze,Jacobs1991-eg}.
 
In the early nineties, Goodale \& Milner argued that, on the basis of neuropsychological, electrophysiological and behavioural evidence, these pathways should be understood as have different goals. The ventral pathway (“vision for perception”) is involved in computing the transformations necessary for the identification and recognition of objects. The dorsal pathway (“vision for action”) is involved in sensorimotor transformations for visually guided actions directed at these objects \cite{Goodale1992-ww}.

It was recently suggested that the brain uses a variety of cost functions for learning \cite{Marblestone2016-od}. These cost functions can be highly diverse. The brain must optimize a wide range of cost functions, such as keeping body temperature constant or optimizing future reward from social interactions. High-level cost functions, by necessity, also shape other cost functions that determine the organization of perception: a cost function that is being optimized to minimize hunger affects the visual recognition cost function as foods have to be recognized. Mechanistically, this could take place directly through, for instance, a reward modulation of object recognition learning, or indirectly through evolutionary pressure on the cost function associated with object recognition learning. In this paper, we try to understand how multiple pathways in the visual cortex might evolve from the perspective of Deep Neural Networks (DNNs) (see \hyperref[sec:box]{box 1}) and cost functions (see \hyperref[sec:box2]{box 2}), and what this implies for how object information is stored in these networks.

We start with a discussion of the relevance of DNNs \cite{LeCun2015-lc,Schmidhuber2015-ae} and, following Marblestone \cite{Marblestone2016-od}, of cost functions for understanding the brain in section \ref{sec:2}. We extend our discussion with the importance of optimizing different cost functions simultaneously, presenting a hypothesis on the relationship between relatedness of tasks and the degree of feature representation sharing.

We test this hypothesis in a computational experiment with DNNs in section \ref{sec:3} to evaluate how much its feature representations contribute to each task. In section \ref{sec:4}, we discuss the degree to which we are able to translate our experimental findings to the division between the ventral and dorsal pathway, the multiple functions of the ventral cortex, and the apparent co-occurrence of both distributed and modular representations related to object recognition.

We finish this paper with a discussion of how this framework can be used experimentally to understand the human brain while elaborating on the limitations of DNNs and cost functions. For brevity, we do not consider models of re-current processing.

%% file: sections/section_2.tex
\section{Multi-task DNNs as models of neural information processing in the brain}
\label{sec:2}
Artificial neural networks are inspired by computational principles of biological neuronal networks and are part of a large class of machine learning models that learn feature representations from data by optimizing a cost function. In this section, we discuss why we believe models based on optimizing cost functions, such as DNNs, are relevant for understanding brain function.

\subsection{Similarities in architecture and behavior between DNNs and the brain}
\label{sec:2_1}
Alexnet \cite{Krizhevsky2012-dk}, a model that is has been used extensively in research relating DNN's to the brain, consists of 7 layers (see \hyperref[sec:box]{box 1}). The first layer consists of filters with small kernels that are applied to each position of the input. In the subsequent four layers this procedure is repeated using the output of the preceding layer. This results in an increase in receptive field (RF) size and concurrently an increase in the specificity of tuning \cite{Zeiler2014-pw}. This increase of receptive field size and tuning specificity traversing the layers resemble the general architecture of feed-forward visual representations in the human brain \cite{Lamme2000-fv,DiCarlo2012-jk}. 

A number of BOLD-MRI studies have revealed that the neural activation's in early areas of visual cortex show the best correspondence with the early layers of DNNs and that higher-tier cortical areas show the best correspondence with higher-tier DNN layers \cite{Guclu2015-gl,Eickenberg2017-mr}. MEG/EEG studies have furthermore shown that early layers of DNNs have a peak explained variance that is earlier than higher-tier DNN layers \cite{Cichy2016-sw,Ramakrishnan2016-da}. In addition, the DNN model has been shown to predict neural responses in IT, both from humans and macaque, much better than any other computational model \cite{Khaligh-Razavi2014-ap,Yamins2014-xi}. 

A number of BOLD-MRI studies have revealed that the neural activations in early areas of the visual cortex show the best correspondence with the early layers of DNNs and that higher-tier cortical areas show the best correspondence with higher-tier DNN layers \cite{Guclu2015-gl,Eickenberg2017-mr}. MEG/EEG studies have furthermore shown that early layers of DNNs have a peak explained variance that is earlier than higher-tier DNN layers \cite{Cichy2016-sw,Ramakrishnan2016-da}. In addition, the DNN model has been shown to predict neural responses in IT, both from humans and macaque, much better than any other computational model \cite{Khaligh-Razavi2014-ap,Yamins2014-xi}.

The correspondence between DNNs and the brain begs the question of the degree to which DNNs show ‘behavior’ similar to humans. Early results indicate that humans and DNNs have a similar pattern of performance in terms of the kinds of variation (size, rotation) that make object recognition harder or simpler \cite{Kheradpisheh2016-ei}. It has also been shown that higher-tier layers of DNNs follow human perceptual shape similarity while the lower-tier layers strictly abide by physical similarity \cite{Kubilius2016-ho}. On the other hand, DNNs are, for instance, much more susceptible to the addition of noise to input images than humans \cite{Jang2017-fw} and the exact degree to which the behavior of DNNs and humans overlap is currently a central topic of research.

As others \cite{Kriegeskorte2015-kw,Yamins2016-cz}, we therefore believe that there is a strong case that DNNs can serve as a model for information processing in the brain. From this perspective, using DNNs to understand the human brain and behavior is similar to using an animal model. Like any model, it is a far cry from a perfect reflection of reality, but it is still useful, with unique possibilities to yield insights in the computations underlying cortical function.

\input{sections/box}

\subsection{Cost functions as a metric to optimize tasks}
\label{sec:2_2}
While deep neural networks offer the representational power to learn features from data, the actual learning process is guided by an objective that quantifies the performance of the model for each input-output pair. Common practice in machine learning is to express such an objective as a cost function \cite{Domingos2012-ex}. As Marblestone and colleagues argue, the human brain can be thought of implementing something very similar to cost functions to quantify the collective performance of neurons and consequently to steer the learning of representations in a direction that improves a global outcome \cite{Marblestone2016-od}.

\subsection{Problem simplification by task decomposition}
\label{sec:2_3}
\begin{figure}[t]
\centering
 \includegraphics[width=0.5\linewidth]{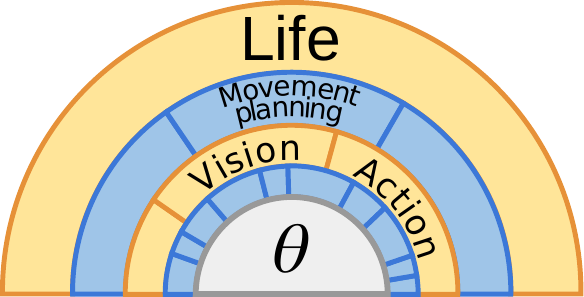}
\caption{\textbf{Hierarchy of tasks related to the objectives the brain has to accomplish.} To make the evolutionary goal of Life tractable, the brain must be able to decompose it into manageable subtasks (blue and yellow arcs). All tasks and their cost functions effectively act on the same set of parameters (gray semicircle), while there may be differing degree of influence.}
\label{fig:1}
\end{figure}

While humans may act under a grand evolutionary objective of staying alive long enough to reproduce, we accomplish many small-scale objectives along the way, like guiding our arms to our mouth to eat or plan our path through the city. Each of these smaller objectives can be thought of as being governed by their own cost functions (see figure \ref{fig:1}). These could be embedded in the brain, either hard coded into the neural substrate by evolution, by sovereign decision making, or as part of meta-learning: learning to learn \cite{Baxter1998-el}. 

While humans may act under a grand evolutionary objective of staying alive long enough to reproduce, we accomplish many small-scale objectives along the way, like guiding our arms to our mouth to eat or plan our path through the city. Each of these smaller objectives can be thought of as being governed by their own cost functions (see figure \ref{fig:1}). These could be embedded in the brain, either hard-coded into the neural substrate by evolution, by sovereign decision making, or as part of meta-learning: learning to learn \cite{Baxter1998-el}.

It has been argued that a task becomes easier to solve if it can be decomposed into simpler tasks \cite{Jacobs1991-eg,Sutton1999-zf}. To support their argument they state that the simple problem of learning the absolute value function can be decomposed into learning two linear functions and a switching function, which leads to a model with fewer parameters that can be trained faster. While such a decomposition could be predefined through the neural substrate, they observe in their experiments that such a decomposition can naturally arise from competitive learning, if the same set of parameters are optimized for multiple tasks. As the decomposition of tasks is underdetermined, the learner may come up with different decompositions, each time it is trained.

The notion of decomposition has been frequently used in machine learning literature on reinforcement learning \cite{Dietterich2000-ka} to increase learning speed and enable the learning of task-local optima that can be reused to learn a superordinate goal. Very often it is even impossible to specify the objective for a complex task so that it is a necessity to decompose it into tractable partial objectives. An example is the objective of vision. Finding an objective for such a broad and vague task appears futile so that it is easier to define a subset of tasks like figure ground segmentation, saliency and boundaries. A noteworthy implementation of such a decomposition is the recent DNN ‘Uber-Net’ \cite{Kokkinos2016-jq}, which solves 7 vision related tasks (boundary, surface normals, saliency, semantic segmentation, semantic boundary and human parts detection) with a single multi-scale DNN network to reduce the memory footprint. It can be assumed that such a multi-task training improves convergence speed and better generalization to unseen data, something that already has been observed on other multi-task setups related to speech processing, vision and maze navigation \cite{Dietterich1990-mx,Dietterich1995-ia,Bilen2016-bv,Mirowski2016-aa,Caruana1998-ix}.

%% file: sections/box.tex
\let\defaultthefigure\thefigure
\renewcommand\thefigure{Box.\arabic{figure}}
\setcounter{figure}{0}

\begin{figure*}

\begin{longfbox}[border-radius=0.5ex, padding={0.5ex,1.5ex}, margin-right=6ex, margin-left=6ex]
\
{Box 1 \rule[-0.2ex]{0.1ex}{1.1em} \textbf{Deep Neural Networks}}
\label{sec:box}

\vspace{-1.0ex}
\noindent\makebox[\linewidth]{\rule{\linewidth}{0.4pt}}
{\small
Artificial neural networks refer to a large class of models loosely inspired by the way brain solves problems with a large number of interconnected units (neurons). The basic computation of a neural network unit is a weighted sum of incoming inputs followed by an activation function i.e a static non-linearity \cite{Rumelhart1988-vw}.
 
Composing a network of many of these basic computational units in more than 3 layers results in what is usually referred to as deep neural network (DNN). While the exact architecture of a DNN varies across applications, the one we are focusing on is the convolutional DNN, specifically designed for inputs with high spatially-local correlation like natural images. Convolution is hereby the process of applying a filter to each position in the image. In the first layer, these filters are able to detect for instance edges and very simple shapes, but composing a hierarchy of these filters allows for great compositional power to express complex features and is an important reason DNNs have proven to be so successful.

As determining these filters by hand is practically impossible DNNs are trained by backpropagation \cite{LeCun1989-cd}, a standard machine learning optimization method based on gradient descent. Given a cost function that determines for an input and an expected output a single error value, backpropagation allows to assign a credit to each single unit in the network to specify how much it contributed to the error. 
 
Recent state-of-the-art neural networks have increased depth, ranging from 16 \cite{Simonyan2014-uo} to 152 \cite{He2015-jc} layers (combined with some architectural advances). While the brain is clearly not shallow, its depth is limited to substantially fewer computational layers considering feed-forward processing \cite{Lamme2000-fv}. However, it has not yet been investigated how the layers of a very deep neural network map to the human brain.

\vspace{0.5em}
\begin{center}
\includegraphics[width=0.9\linewidth]{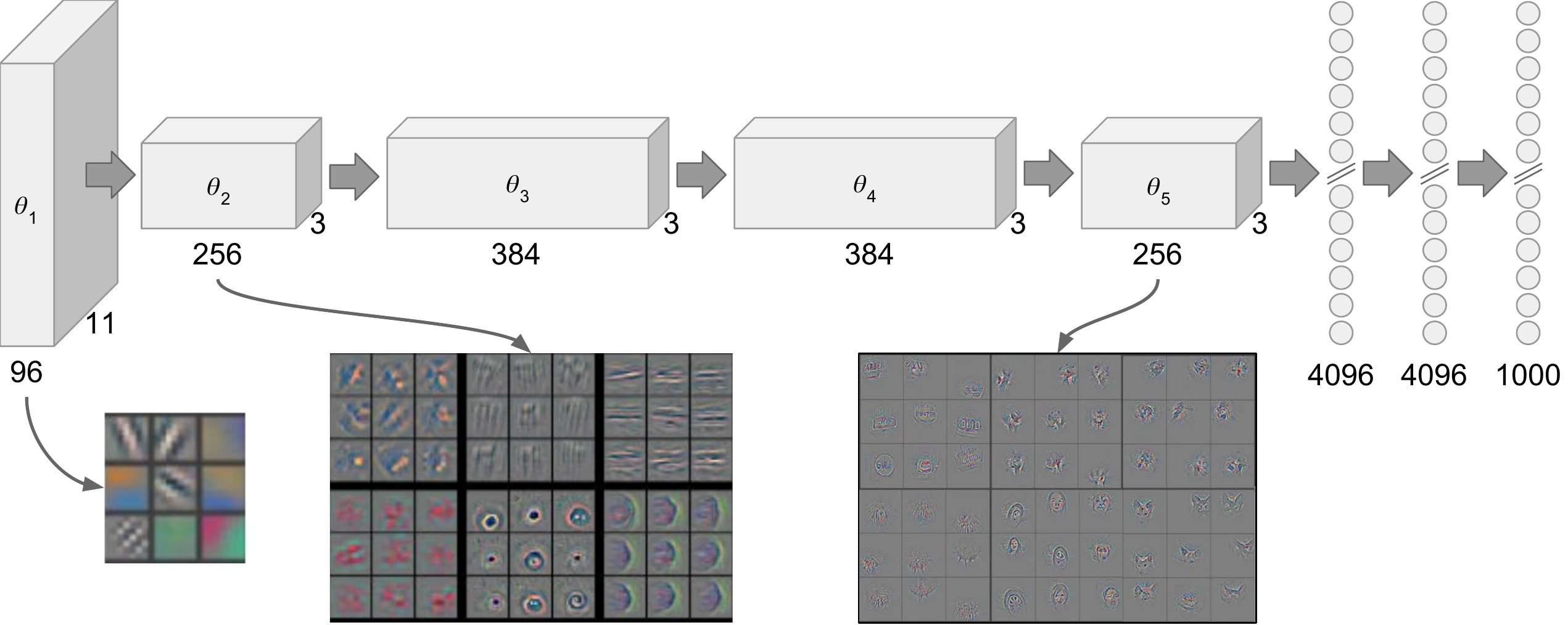}
\caption{\footnotesize\textbf{Essential architecture of DNN AlexNet and filter visualization.} AlexNet consists of 5 convolutional layers represented by boxes and 3 fully connected layers of which the last is the output layer with 1000 units. The number of filters in a layer as well as the filter dimension is noted under each box. Below are selected filters visualized to show the increasing complexity of features they represent (adopted from Zeiler et al 2014).}
\label{fig:alexnet}
\end{center}
}

{Box 2 \rule[-0.2ex]{0.1ex}{1.1em} \textbf{Cost Functions}}
\label{sec:box2}

\vspace{-1.0ex}
\noindent\makebox[\linewidth]{\rule{\linewidth}{0.4pt}}
{\small
A cost function maps a set of observable variables to a real value representing the `loss’ of the system. 
Optimization then aims to minimize this loss, for instance by changing tunable parameters $\theta$ in the system. 
While a cost function is defined as the composition of mathematical operations, e.g. mean squared error, we expand the definition here to include the set of observed variables.
This allows us to regard two cost functions composed of the same mathematical operations as distinct when the set of observed targets, in order to solve two different objectives, is different.

For a predictive brain in a moving organism, the system tries to optimize actions, and sequences thereof that minimize one or more cost functions; these actions in turn are specified by a plethora of parameters, like synaptic efficacies and hormone levels. 
It is these parameters that are adjusted to change the actions that the system takes in a given environment to decrease the cost.  
 
Mathematically, we can specify the collective sensory input into the brain at any point in time as $S$, and the joint output of muscle tensions as $O$. A cost function maps the outputs $O$ into a value, $f(O)$, that is minimized by adjusting the parameters $\theta$: learning. Multiple cost functions arise naturally when different measured quantities are to be optimized: if $t=f_{thirst}(O, \Theta)$ corresponds to the degree of thirst, and the system also has to optimize financial welfare $d=f_{fw}(O,\Theta)$, the system has to find the optimum values of theta that maximize both functions. We can jointly optimize these two cost functions by specifying a single combined cost function:  $G = f_{fw}(O,\theta) + \lambda f_{thirst}(O,\theta)$, where $\lambda$ is a weight that measures the relative importance of the two cost functions. Such joint cost functions can be learned with a single network, where the degree to which shared representations (in the form of shared learned features) help or hurt with the optimization task is variable \cite{Caruana1998-ix}. The shape of the cost has likely evolved such that they help make most sense of our environment \cite{Marblestone2016-od}: a loss may measure the absolute deviation from some target value, or the square of this difference, or any other mapping. 
}
\end{longfbox}
\end{figure*}

\renewcommand\thefigure\defaultthefigure
\setcounter{figure}{0}

%% file: sections/section_3.tex
\section{Functional organization in multi-task DNNs}
\label{sec:3}
One hypothesis for the emergence of different functional pathways in the visual system is that learning and development in the cortex is under pressure of multiple cost functions induced by different objectives. It has been argued that the brain can recruit local populations of neurons to assign local cost functions that enable fast updating of these neurons \cite{Marblestone2016-od}.
We explore in this section the ramifications of multiple cost functions acting on the same neurons by translating the problem to instances of multi-task DNNs sharing the same parameters. By observing the contributions each feature representation in a DNN has to each task, we will draw conclusions about the functional separation we observe in the visual cortex in section \ref{sec:4}.

\subsection{Hypothesis}
\label{sec:3_1}
\begin{figure*}[t]
\centering
\includegraphics[trim=15mm 90mm 15mm 0mm, clip, width=1\linewidth]{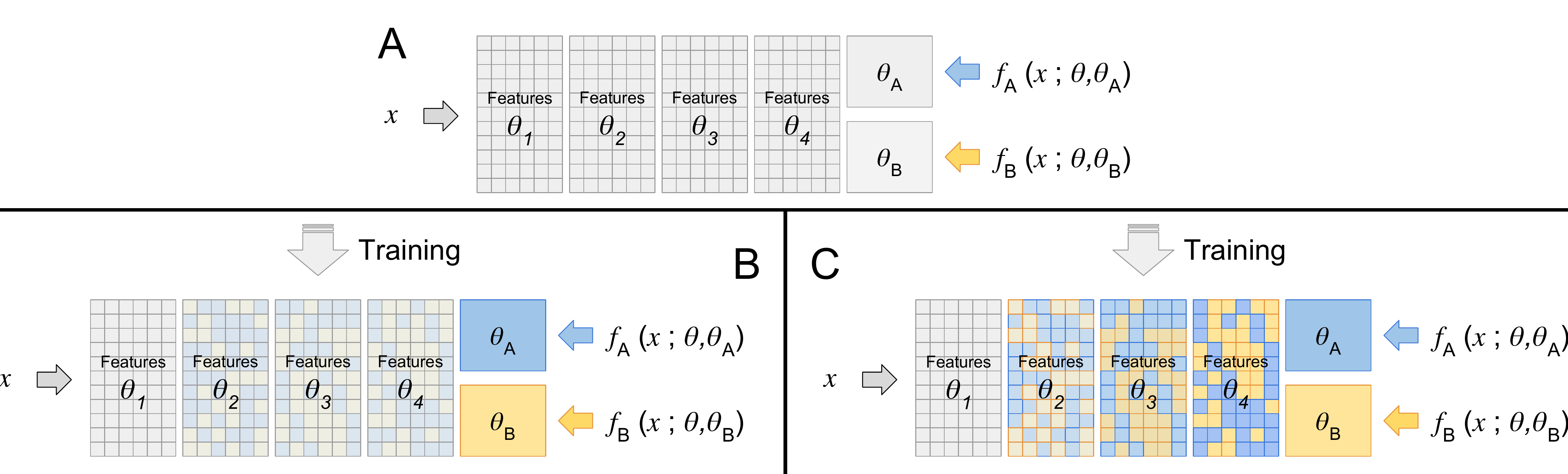}
\caption{\textbf{Task relatedness and feature representation sharing in deep neural networks.} Given a multi-layered neural network with a set of feature representations $\theta$ (indicated by cells) that optimize differently related tasks, we conjecture that the degree to which representations can be shared is dependent on the generalizability, which reduces with the depth of the network for single modality inputs. The generalizability is indicated by the strength of the color. Gray tones indicate high generalizability, while strong colors indicate features that are tuned to one respective cost function. \textbf{A} | Initial, untrained network configuration with 5 layers for a single modality input $x$. Cost functions $f_A$ and $f_B$ have direct access to their respective parameters $\theta_A$ and $\theta_B$. \textbf{B} | Two strongly related tasks inducing features that are generalizable to both tasks. Little function-specificity identifiable. \textbf{C} | Two largely unrelated tasks. While early simple features representations can be shared, intermediate and higher level representations are likely to be exclusive to their respective cost function due to their task-specificity.}
\label{fig:2}
\end{figure*}

Given two cost functions that optimized two related tasks, which both put pressure on the same set of parameters, we conjecture that the parameters learned will be general enough to be used for both tasks (see figure \ref{fig:2}B). In contrast, we speculate that, when the tasks are unrelated, two subsets of parameters will emerge during learning that each lie within their task-respective feature domain (see figure \ref{fig:2}C). Because the amount of feature representation sharing is determined by the relation between tasks, and ultimately by the statistics of the credit assignments, we predict an upper to lower tier gradient of feature representation sharing with the least sharing in higher tier layers.

\subsection{Training models for multiple tasks}
We test this hypothesis on feature representation sharing with DNNs trained for two tasks simultaneously. We construct two example setups involving a pair of related tasks (which we call RelNN), namely the simultaneous classification of ordinate and subordinate categories of objects in images, and a pair of unrelated tasks (which we call UnrelNN) namely the classification of objects and text labels in images (see figure \ref{fig:3}). As the relatedness of tasks is not clearly defined and an open problem \cite{Caruana1998-ix,Zhang2014-ta}, the tasks were selected based on the assumption that text recognition in UnrelNN is mostly independent of object recognition while in contrast ordinate level classification in RelNN is highly dependent on the feature representations formed for subordinate level classification.

\subsubsection{Training setup}
\begin{figure}[t]
\centering
\subfloat[]{
 \includegraphics[height=0.35\linewidth]{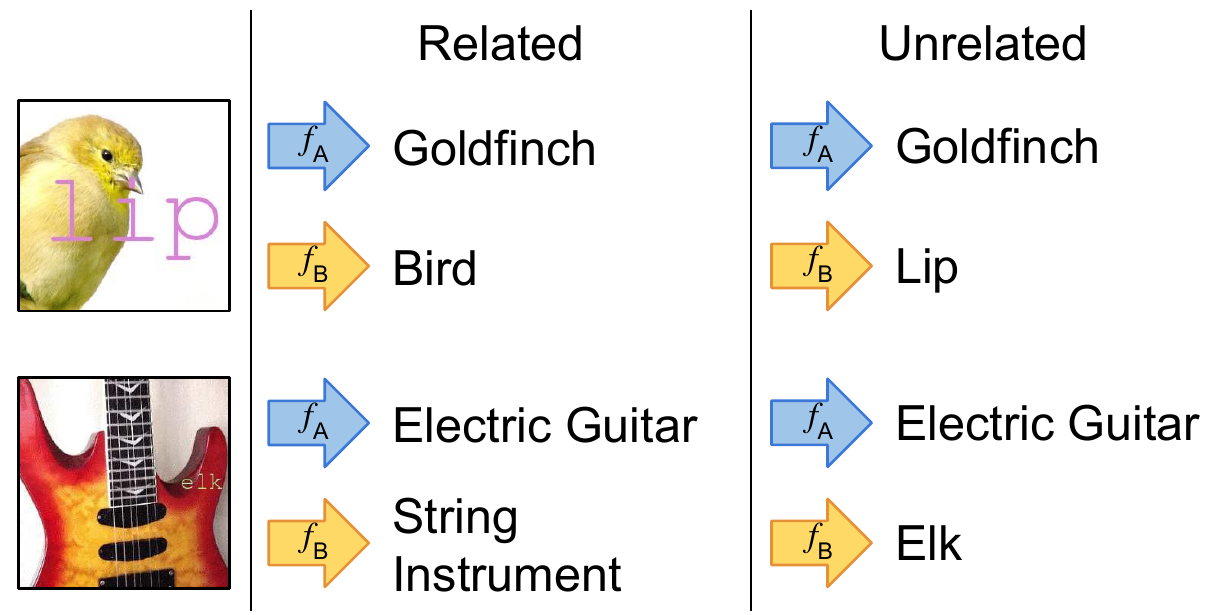}
 \label{fig:3_1}
}
\subfloat[]{
 \includegraphics[height=0.35\linewidth]{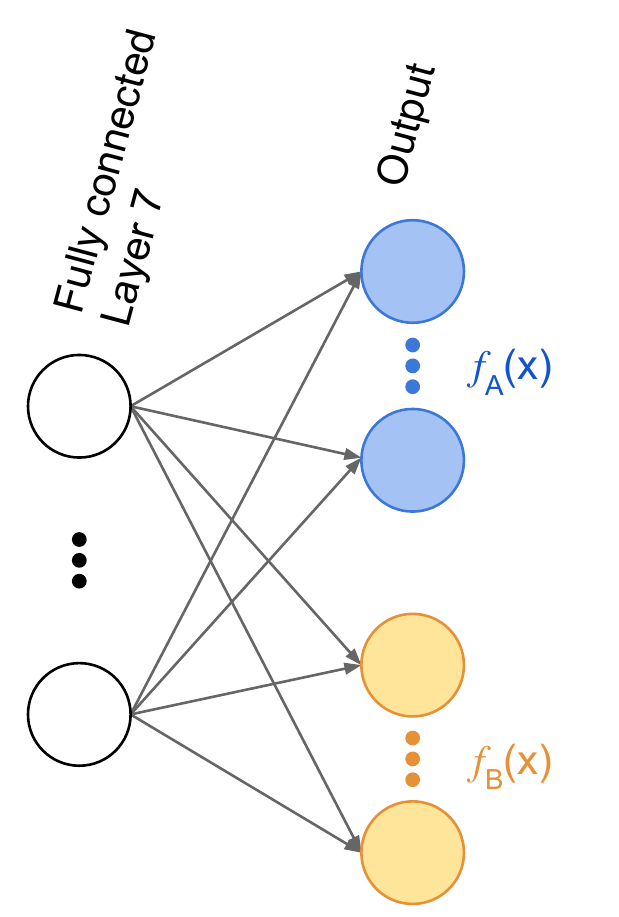}
 \label{fig:3_2}
}
\caption{\textbf{Multi-Task-Learning setup.} (a) Two example images and their corresponding classification for our example setups of related and unrelated tasks. (b) To classify an input into two categories from different domains using AlexNet, the output layer is split in two where each split has its own softmax activation layer.}
\label{fig:3}
\end{figure}

Both setups were implemented by training a version of AlexNet \cite{Krizhevsky2012-dk} on approximately half a million images from the ImageNet database \cite{Russakovsky2015-js} each
\footnote{The code, data and pretrained models are available here: \\\url{https://github.com/mlosch/FeatureSharing}}. 
To optimize the models for two tasks simultaneously, the output layer of AlexNet was split into two independent layers. 
Both models were trained on an identical set of images consisting of 15 ordinate classes further divided into 234 subordinate classes, each image augmented with an overlay of 3 letter labels from 15 different classes (see figure \ref{fig:3_1}). 
The overlays were randomly scaled, colored and positioned while ensuring that the text is contained within the image boundaries. 
Furthermore to enable the networks to classify two tasks at once, the output layer was split in two independent layers (see figure \ref{fig:3_2}) for which each had its own softmax activation. 
For classification performance results see table \ref{tab:1}.

\begin{table}[t]
    \begin{center}
    \begin{tabular}{l|c|c|}
        \cline{2-3}
             & \multicolumn{2}{c|}{Top-5-error} \\
         & \begin{tabular}[c]{@{}c@{}}Subordinate-level\\recognition\end{tabular} & \begin{tabular}[c]{@{}c@{}}Ordinate-level/Text\\recognition\end{tabular} \\ \hline\hline
        Chance & 97.9\% & 66.7\% \\ \hline
        RelNN & 14.0\% & 2.9\% \\ \hline
        UnrelNN & 15.2\% & 4.9\% \\ \hline
    \end{tabular}
    \end{center}
    \caption{\textbf{Classification errors.} Comparison of the error rates of RelNN and UnrelNN on a validation set of 11,800 images. The Top-5-error is defined as the correct prediction not being under the 5 most likely predictions. Both models were trained for $90$ epochs until convergence with Nesterov accelerated gradient descent \protect\cite{Nesterov1983-dp} with momentum of $0.9$, starting with a learning rate of $0.01$ and decreasing it every $30$ epochs by a factor of $10$.}
    \label{tab:1}
\end{table}

\subsubsection{Measuring feature representation contribution}
To determine the degree of feature representation sharing in a neural network we measure the contribution each feature representation has to both tasks. Our method is inspired by the attribute contribution decomposition by \cite{Robnik-Sikonja2008-of} which has recently been used to visualize the inner workings of deep convolutional networks \cite{Zintgraf2016-yc}. The method is used to marginalize out features in the input image in the shape of small image patches, to observe the impact on the classification. In comparison, our method considers feature representations instead of features as we are not interested in the contribution of particular feature instances. The interested reader is referred to appendix \ref{sec:appendix} for the definition and derivation of the task contribution.

\subsubsection{Results}
\begin{figure*}[!htb]
\centering
\subfloat{
 \includegraphics[width=1\linewidth]{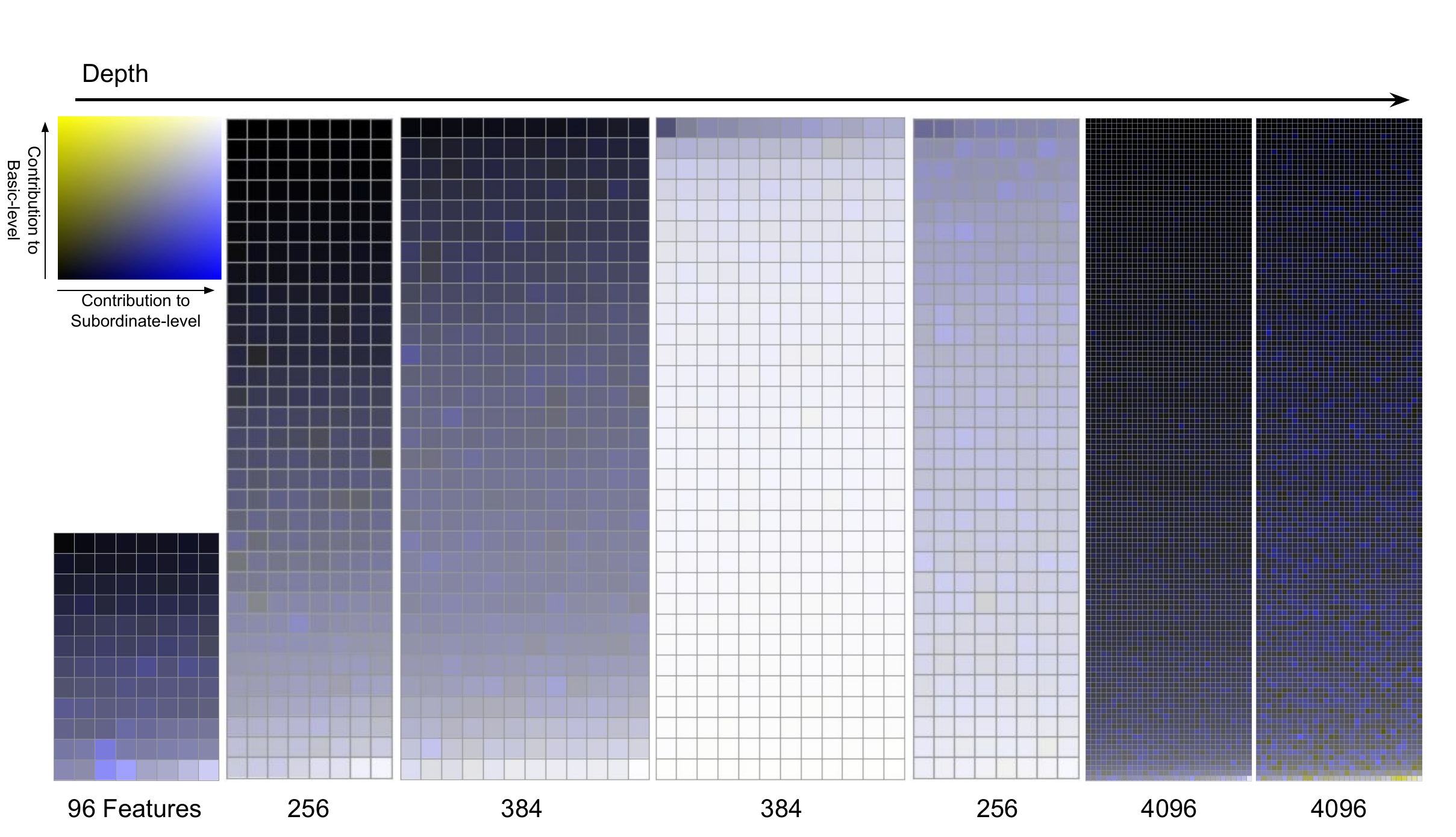}
 \label{fig:4_1}
}
\\
\subfloat{
 \includegraphics[width=1\linewidth]{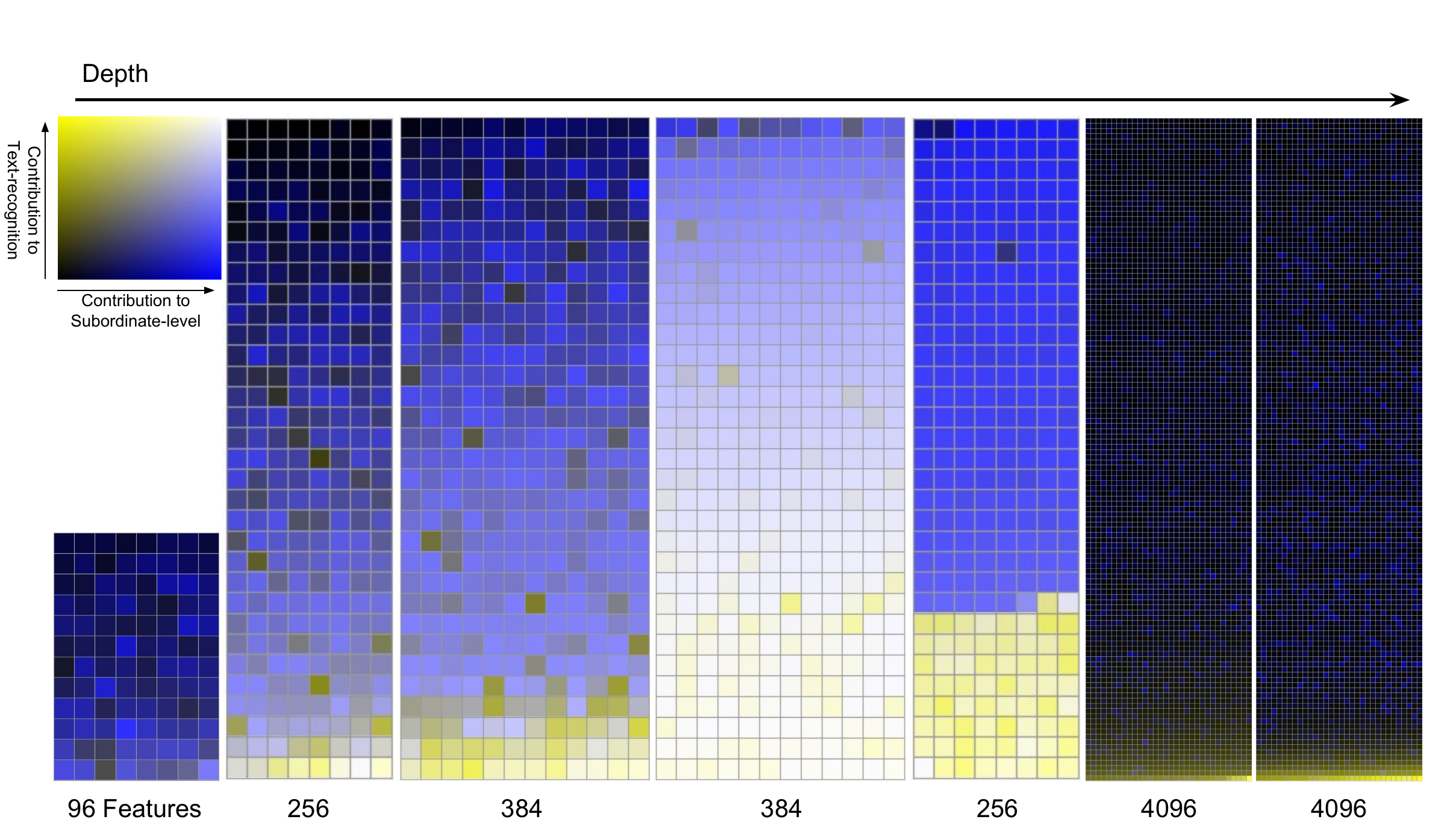}
 \label{fig:4_2}
}
\caption{\textbf{Composition of feature representation contribution in DNNs to dual task.} (Best viewed in color) Each cell represents a feature representation in a neural network and its contribution. Task description and color coding of the contributions are displayed in the top left corner of each visualization. The cells are ordered by contribution magnitude of the yellow task so that the first cell in each layer displays the representation that contributes the least.  (a) Contributions to RelNN, subordinate- and ordinate-level-recognition. (b) Contributions to UnrelNN, subordinate-level- and text-recognition.}
\label{fig:4}
\end{figure*}

We visualize the layer-wise task contributions by unrolling the feature representations of a layer on a rectangle and coloring each resulting cell by the composition of its contribution. Blue is used as indicator for the subordinate-level recognition task and yellow as indicator for the text- and basic-level-recognition task respectively. Equal contribution to both tasks results in grayish to white tones while little contribution to either task causes dark to black tones (see figure \ref{fig:4} for the color coding). A high degree of feature representation sharing would hereby generate cells colored in the range from black and gray to white, while low degree of sharing would result in more pronounced and clearly distinguishable colors of yellow and blue.
 
The two visualizations in figure \ref{fig:4} show a substantial difference in feature representation contribution as the representations in layer 2 to 5 of the RelNN contribute to both tasks much more equally than the representations of the UnrelNN. This is in line with our expectation depicted in figure \ref{fig:2} and our choice of setups. Contrary to our prediction, the degree of feature representation sharing in layer 1 of the UnrelNN is lower than expected; this can be explained by assuming that text recognition is mostly independent of all features but horizontal and vertical lines. Note also that most of the representations in the fully connected layers in both setups have only little contribution. This might seem counter-intuitive at first sight but is an effect of the abundance of representations coupled with the training scheme involving dropout. Dropout significantly reduces co-dependencies between units \cite{Dahl2013-io} resulting in only small changes in classification probability after marginalizing out a single representation.
 
We also observe that there is a dominance of blue cells expressing low contribution to the text- and basic-level-recognition task but high contribution to the subordinate-level-recognition task. We conjecture that this is because the subordinate-level-recognition task uses a larger fraction of units to distinguish between 200 classes.
 
Comparing the layers of both networks, it becomes evident that there generally is a higher degree of feature representation sharing in the RelNN consistent with the idea that relatedness between tasks and therefore cost functions strongly influences the degree of feature representation sharing across layers. More importantly, these results demonstrate that these types of ideas can be translated, using the right image data-sets and task-labels, into quantifiable predictions on the degree of feature sharing that might be observed in the brain.

%% file: sections/section_4.tex
\section{Implications of models optimized for multiple tasks for understanding the visual system}
\label{sec:4}
In section \ref{sec:3} we presented an example in which the degree to which feature representations can be shared in a neural network depended on the relatedness of the tasks they are optimized for. In a neural population under pressure of the optimization for two unrelated tasks and the pressure to optimize the length of neuronal wiring \cite{Chklovskii2004-ki}, a spatial segregation is likely to occur, resulting in anatomically and functionally separate pathways. In this section we consider to what degree we can understand the organization of the visual system from the perspective of a DNN that has been trained on multiple tasks and discuss three hypotheses derived from the simulations.

\subsection{The visual system optimizes two cost functions of unrelated tasks}
\label{sec:4_1}
The early visual cortex has neurons that respond to properties such as orientation, wavelength, contrast, disparity and movement direction that are relevant for a broad range of visual tasks \cite{Wandell1995-yx}. Moving upwards from early cortex we see a gradual increase in the tuning specificity of neurons resulting in the dorsal and ventral pathways that have, as has become clear the last 25 years, unrelated goals \cite{Goodale1992-ww}. The dorsal pathway renders the representation of objects invariant to eye-centered transformations in a range of reference frames to allow efficient motor planning and control \cite{Kakei1999-vh}, while the ventral pathway harbors object-centered, transformation invariant features \cite{Leibo2015-nb,Higgins2016-aj} to allow efficient object recognition. 
 
These observations concur well with the predictions and experimental results we made about feature representation sharing in DNNs. Given that the two tasks, vision for recognition and vision for action, are mostly unrelated we can understand the gradual emergence of functional and anatomical separation between these systems from this perspective.
 
Nonetheless, we note that the functional units of the pathways beyond the occipital lobe are not entirely separated and cross-talk does exist between these pathways \cite{McIntosh2009-ua,Farivar2009-qe,De_Haan2011-aq,Van_Polanen2015-tg}: a phenomenon we also observed in our experiment in section \ref{sec:3}. In the UnrelNN, there are feature representations that contribute to both tasks throughout all layers of the network. Consequently the brain might trade off contribution and wiring length so that neurons that contribute little are tolerable to have long wiring to the functional epicentre.
 
As a whole the existence of two pathways guided by two cost functions of unrelated tasks might be seen as an illustration of the efficient decomposition of the overall vision function.

\subsection{The visual pathways contain further task decompositions each with their own cost functions}
\label{sec:4_2}
\begin{figure*}[t!]
\centering
\includegraphics[width=1\linewidth]{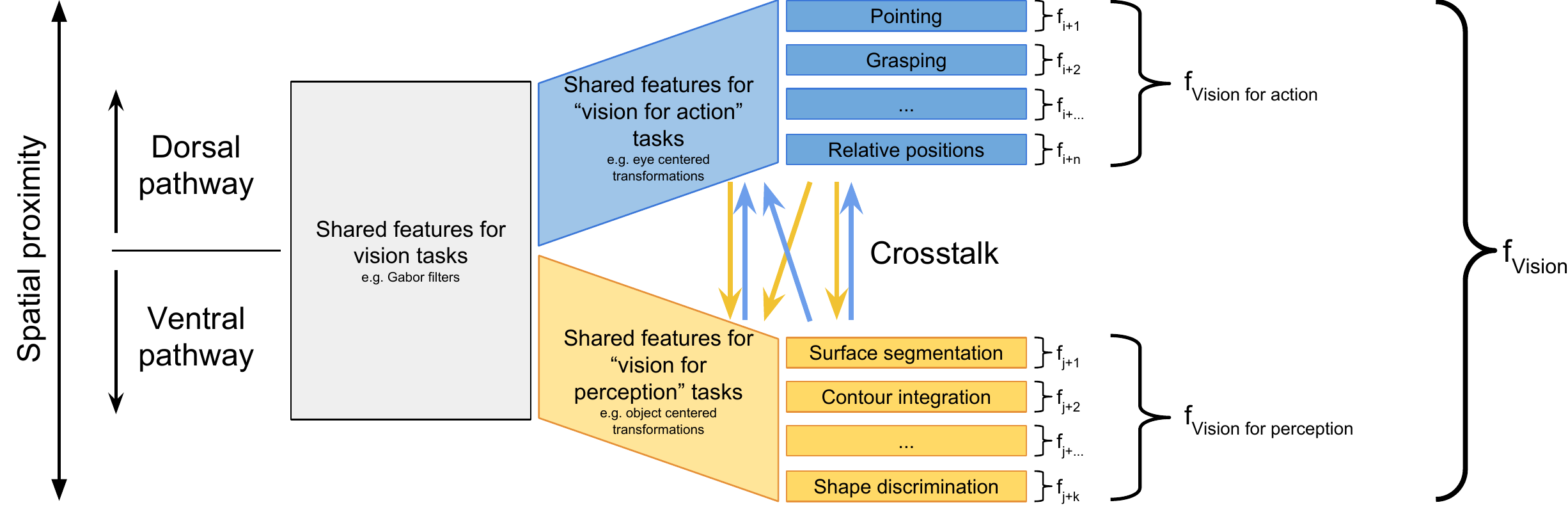}
\caption{\textbf{How functional pathways in the visual system could be associated with cost functions.} (Vision for perception pathway in blue, vision for action pathway in yellow). Within the pathways are streams that develop under guidance from cost functions which are a direct decomposition of the pathways cost function. Feature representations that are learned for one task can still be used by units in the other pathway (crosstalk arrows). Both pathways share the same input units that either develop through the relation between tasks and/or evolutionary or developmental learning.}
\label{fig:5}
\end{figure*}

We further generalize our perspective on cost function optimization of the visual system via the general observation made from machine learning that a complex task becomes simpler to solve if it is decomposed into simpler smaller tasks (see section \ref{sec:2_3}). Given that the tasks we assign to the visual pathways are rather complex and vague we conjecture that there might be a broad range of cost functions active in the pathway regions to optimally decompose the task of vision resulting in a schematic similar to figure \ref{fig:5}.

The ventral and dorsal pathways are each involved in a multitude of tasks serving the overall goals of vision for perception and vision for action. Examples of subordinate tasks for vision for action are localization, distance, relative position, position in egocentric space and motion and these interact with the goals that are part of vision for action: pointing, grasping, self-termination movements, saccades and smooth pursuit \cite{De_Haan2011-aq}. Sub-ordinate tasks for vision for perception include contour integration, processing of surface properties, shape discrimination, surface depth and surface segmentation. These in turn interact with executing the goals that are part of vision for perception: categorization and identification of object but also scene understanding \cite{Groen2017-ax}.

Reasoning from this framework we can also understand the existence of multiple ‘processing streams’ within the dual pathways. For instance, within ventral cortex there appears to be a pathway for object recognition and a pathway for scene perception. The object recognition pathway consists of areas like V4 which responds to simple geometric shapes and the anterior part of inferior temporal (aIT) that is sensitive for complete objects \cite{Kravitz2013-rv}. The scene recognition pathway contains areas such as the occipital place area (OPA), involved in the analyses of local scene elements and the parahippocampal place area (PPA) which responds to configurations of these elements \cite{Kamps2016-ov}. The tasks of scene and object perception are closely related; scenes consist of objects. However, scene perception involves relating the positions of multiple objects to each other, scene gist and navigability \cite{Groen2017-ax}. From our framework we would predict that an area like OPA is mainly involved in the task of scene perception but has RFs that are also used for object perception and the opposite pattern for V4. Crucially, we believe this framework can be used to generate quantitative predictions for this amount of sharing.

\subsection{Distributed versus modal representations}
\label{sec:4_3}
How information is represented is one of the major questions in cognitive neuroscience. When considering object based representations both distributed \cite{Haxby2001-qj,Avidan2009-cu} and module-based representations \cite{Cohen2000-bs,Kanwisher2000-ei,Puce1995-qs} have been observed.
 
Module-based representations, and theories stressing their importance, point to the existence of distinct cortical modules specialized for the recognition of particular classes such as words, faces and body parts. These modules encompass different cortical areas and, in case of the fusiform face area and visual word form area, even similar areas but in different hemispheres \cite{Plaut2011-gq}. Conversely, distributed theories of object recognition point to the possibility to decode information from a multitude of classes from the patterns of activity present in a range of cortical regions \cite{Haxby2001-qj,Avidan2009-cu}.
 
If we consider feature representations in the early and intermediate layers of the UnrelNN (figure \ref{fig:4}) as a reasonable approximation of representations in early / intermediate visual areas, we note that most units are being shared by both streams. However, some units contribute more to one than the other task and are spatially intermingled at the same time. An external observer, analyzing the activity of these representations under stimulation with pattern analysis would conclude that information from both tasks is present, and conclude that a distributed code is present. If the same observer would investigate the representations at the top of the stream the observer would conclude that there is an area dedicated to the analysis of text and another to the analysis of the subordinate task. 
 
Translated to the visual system this would mean that distributed representations should be observed in areas such as posterior inferior temporal (pIT), OPA and V4 because these units are activated by multiple tasks but with a different weighting. Vice versa, at the top of a pathway or stream the network would show a strong module based pattern of activation. In sum, multi-task DNNs provide a framework in which we can potentially understand that both modal and distributed representations can be observed experimentally but suggest that the patterns of activity should be interpreted as emerging from the network as a whole.

%% file: sections/section_5.tex
\section{Discussion}
\label{sec:5}
Following Marblestone and colleagues \cite{Marblestone2016-od}, and the strength of the similarities between DNNs and the visual brain, we hypothesize that cost functions, associated with different tasks, are a major driving force for the emergence of different pathways.
 
A central insight from machine learning is that functions become easier to learn when they are decomposed as a set of unrelated subtasks. As a whole, the existence of two pathways guided by two cost functions of unrelated tasks might be seen as an illustration of the efficient decomposition of the overall vision function \cite{Sutton1999-zf}. Observing that DNNs decompose a problem in multiple steps, with the earlier layers related to the input and later layers related to outputs demanded for the task, we hypothesized that the degree of feature representation sharing between tasks, will be determined by the relatedness of the tasks with an upper-to-lower tier gradient.
 
On this basis, we performed simulations that confirm that units in a DNN show a strong degree of sharing when tasks are strongly related and a separation between units when tasks are unrelated. The degree to which this framework will be useful depends on the degree to which understanding elements of brain function using DNNs is valid which is discussed in section \ref{sec:5_1} and \ref{sec:5_2}. 
Subsequently, we will argue that having multiple pathways within a multi-task network might also help explaining catastrophic forgetting, the phenomenon that an old task is overwritten by learning a new task (section \ref{sec:5_3}). Next, we will discuss the ‘vision for perception’ and ‘vision for action’ framework (section \ref{sec:5_4}), and finally we discuss the possibilities of using multi-task for further understanding the brain and ways in which our current analysis approach can be extended (section \ref{sec:5_5}).
 
\subsection{The biological realism of machine learning mechanisms}
\label{sec:5_1}
While there has been much progress in the field of Deep Learning, it remains a question how and if the weights of neurons are updated in learning under the supervision of cost functions in the brain, that is, what the actual learning rules of the brain are.
 
DNNs are trained using back-propagation, an algorithm believed to miss a basis in biology \cite{Crick1989-dl,Stork1989-kg}. Some of the criticisms include the use in backpropagation of symmetrical weight for the forward inference and backward error propagation phase, the relative paucity of supervised signals and the clear and strong unsupervised basis of much learning. Recent research has shown that the symmetrical weight requirement is not a specific requirement \cite{Lillicrap2016-eg}. Roelfsema \& Van Ooyen already showed in \cite{Roelfsema2005-so} that a activation feedback combined with a broadly distributed, dopamine-like error-difference signal can on average learn error-backpropagation in a reinforcement learning setting. Alternative learning schemes, like Equilibrium Propagation \cite{Scellier2017-ai} have also been shown to approximate error-backpropagation while effectively implementing basic STDP rules.
 
Alternatively, effective deep neural networks could be learned through combination of efficient unsupervised discovery of structure and reinforcement learning. Recent work on predictive coding suggests this might indeed be feasible \cite{Whittington2017-if}. Still, the learning rules that underpin deep learning in biological systems are very much an open issue.

\subsection{Cost functions as the main driver of functional organization}
\label{sec:5_2}
Reviewing literature on the computational perspective for functional regions in the visual system, we conclude that each region might be ultimately traced back to being under the influence of some cost function that the brain optimizes and its interplay or competition for neurons \cite{Jacobs1991-eg} with other cost functions resulting in different degrees of feature representation sharing. The domain-specific regions in the ventral stream for example may be caused by a cost function defined to optimize for invariance towards class-specific transformations \cite{Leibo2015-nb}, of which the Fusiform Face Area could additionally be bootstrapped from a rudimentary objective, hard coded by genetics, to detect the pattern of two dots over a line \cite{McKone2012-bv,Marblestone2016-od}. Finally, as we argued in section \ref{sec:4}, the functional separation of the ventral and dorsal pathway can be associated with two cost functions as well. We emphasize that the precise implementation of these cost functions is unknown and note the concept of the task ``vision for recognition" and ``vision for action" is merely a summary of all the subordinate tasks that these two tasks have been decomposed into, as argued in section \ref{sec:2_3} and the \hyperref[sec:box]{cost function box}.

Reviewing literature on the computational perspective for functional regions in the visual system, we conclude that each region might be ultimately traced back to being under the influence of some cost function that the brain optimizes and its interplay or competition for neurons \cite{Jacobs1991-eg} with other cost functions resulting in different degrees of feature representation sharing. The domain-specific regions in the ventral stream for example may be caused by a cost function defined to optimize for invariance towards class-specific transformations \cite{Leibo2015-nb}, of which the Fusiform Face Area could additionally be bootstrapped from a rudimentary objective, hard coded by genetics, to detect the pattern of two dots over a line – being the basic constellation of a face \cite{McKone2012-bv,Marblestone2016-od}. Finally, as we argued in section \ref{sec:4}, the functional separation of the ventral and dorsal pathway can be associated with two cost functions as well. We emphasize that the precise implementation of these cost functions is unknown and note the concept of the task “vision for recognition” and “vision for action” is merely a summary of all the subordinate tasks that these two tasks have been decomposed into, as argued in section \ref{sec:2_3} and the \hyperref[sec:box]{cost function box}.
 
\subsection{Multiple pathways as a solution for catastrophic forgetting}
\label{sec:5_3}
While joint cost functions can be learned when the quantities needed by the cost functions are all present at the same time, most animals are continually learning and different aspects of cost functions are present at different times. Then, it is well known that standard neural networks have great difficulty learning a new task without forgetting an old task, so-called catastrophic forgetting. Effectively, when training the network for the new task, the parameters that are important for the old task are changed as well, with negative results. While very low learning rates, in combination with an alternating learning scheme, can mitigate this problem to some degree, this is costly in terms of learning time. For essentially unmixed outputs, like controlling body temperature and optimizing financial welfare, an easy solution is to avoid shared parameters, resulting in separate neural networks, or “streams”. Similarly, various properties can be derived from a single stream, like visual aspects (depth, figure-ground separation, segmentation), from an object recognition stream, where each aspect substream is learned via a separate cost function. For tasks sharing outputs, and thus having overlap over different tasks, evidence increasingly suggests that the brain selectively “protects” synapses for modification by new tasks, effectively “unsharing” these parameters between tasks \cite{Kirkpatrick2016-qa}.
 
\subsection{What and where vs. vision for action and perception}
\label{sec:5_4}
Goodale \& Milner argued that the concept of a ‘what and where’ pathway should be replaced by the idea that there are two pathways with different computational goals, vision for perception and vision for action, summarized as a ‘what’ and ‘how’ pathway \cite{Goodale1992-ww}. Insights from the last 25 years of research in vision science have shown that the original idea of a what and where pathway lack explanatory power. It is clear that RFs in inferior temporal cortex are large when objects are presented on a blank background \cite{Gross1985-hg}. However, these become substantially smaller and thereby implicitly contain positional information, when measured against a natural scene background \cite{Rolls2003-aq}. Interestingly, studies on DNNs have shown that approximate object localization can be inferred from a CNN trained on only classification, although the spatial extend of an object cannot not be estimated \cite{Oquab2015-lj}.
 
With regards to the dorsal pathways it has been observed that there are cells relating to gripping an object that are specific for object-classes \cite{Brochier2007-bk} showing that this pathway contains, in addition to positional information, categorical information. These observations are in direct opposition to one of the central assumptions, a strong separation between identity and location processing, of the ‘what’ and ‘where’ hypothesis. It is now abundantly clear that the move from ‘what’ and ‘where’ pathway to ‘what’ and ‘how’ pathways and moving from input to function fits particularly well with vision as a multi-task DNN.

\subsection{Future research}
\label{sec:5_5}
Originally DNNs were criticised for being “black” boxes, and using DNNs to understand the brain would equate to replacing one black box with another. Recent years have shown a rapid increase in our understanding of what makes a DNN work \cite{LeCun2015-lc,Simonyan2014-uo,Zeiler2014-pw}  and how to visualize the features \cite{Zintgraf2016-yc,Zhou2014-ue,Zeiler2014-pw} that give DNNs its power.
 
These developments illustrate that DNNs are rapidly becoming more “gray” boxes, and are therefore a promising avenue into increasing our understanding of the architecture and computations used by the visual system and brain.
 
We therefore believe it is sensible to investigate to which degree multi-task DNNs, trained using the same input, will allow us to understand the functional organisation of the visual system. Using the analytical framework introduced in section \ref{sec:3}, we can generate a fingerprint for each of the layers in a network based on the degree of feature representation sharing. This can be subsequently related to the activation patterns, evoked by different tasks observed within different cortical areas. Alternatively it is possible to compare representational dissimilarity matrices \cite{Kriegeskorte2008-vo} obtained from single and multitask-DNNs and determine which better explain RDMs obtained from cortical areas.
 
An open question remains how subtasks and their associated cost functions are learned from overall goals/general cost functions, both in machine learning \cite{Lakshminarayanan2016-ro} and in neuroscience \cite{Marblestone2016-od,Botvinick2009-qs}.

%% file: sections/appendix.tex
\clearpage
\begin{appendix}
\section{Measuring parameter contribution}
\label{sec:appendix}

\subsection{Marginalization of parameters}
In models that are able to handle the lack of information about a particular representation like in na\"ive Bayesian classifiers, the contribution can be measured by marking the representation as unknown. Typically though, neural networks are not able to handle missing information and setting the parameters of a representation to zero will still result in interpretable information for subsequent layers. While removing a feature representation and retraining the network would alleviate this issue, quantifying the contribution of thousands of representations this way is generally unfeasible. Instead we make use of the models classification probabilities given by the softmax activation output which allows us to estimate the classification probability while lacking a representation by marginalizing it out via standard method from statistics. Marginalization effectively computes the weighted average of the classification probabilities after the representation has been replaced with random values sampled from an appropriate distribution. See equation \ref{eq:1} for the mathematical definition used for our evaluation.
\begin{align}
    p(y|x,\Theta_{\setminus \theta}) = \sum_{\theta} p(y | x,\Theta) p(\theta)
    \label{eq:1}
\end{align}
$p(y|x,\Theta)$ defines here the probability of input $x$ belonging to class $y$ and $p(y|x,\Theta_{\setminus \theta})$ the probability if $\theta$ is unknown. Note that a feature representation is represented by its parameters $\theta$, which in turn consists classically of a weight $w$ and a potential bias $b$ in a neural network setting. $\Theta$ defines then the set of all parameters such that $\theta \in \Theta$. Each classification probability is eventually weighted by the prior probability of the sample $\theta$ expressing the likelihood the parameter in question takes value $\theta$. We used 100 samples in our experiments to approximate the contribution.

\subsubsection{Derivation}
Given a parametric model like a DNN that is described by its parameters $\Theta$, we can express the probability of input $x$ belonging to class $y$ as $p(y|x,\Theta)$, where the probabilities are given by the softmax output layer. To measure the contribution of a feature generated by parameter $\theta \in \Theta$, we are interested in what the probability is when $\theta$ is missing or unknown. By assuming that the input is independent of the parameters as well as the parameters are independent of each other, such that $p(x,\Theta) = p(x)p(\Theta)$ and $p(\Theta) = p(\Theta_{\setminus \theta}) p(\theta)$ and by treating the parameters as random variables we can marginalize out $\theta$ as follows.
\begin{align} 
p(y|x,\Theta_{\setminus \theta}) &= \frac{\int_{\theta} p(y,x,\Theta) d\theta}{\int_{\theta} p(x, \Theta) d\theta}\\ 
&= \frac{\int_{\theta} p(y | x,\Theta) p(x,\Theta_{\setminus \theta}) p(\theta) d\theta}{p(x, \Theta_{\setminus\theta}) \int_{\theta} p(\theta) d\theta} \\ 
&= \int_{\theta} p(y | x,\Theta) p(\theta) d\theta 
\end{align}
As the integral over all possible values of $\theta$ is intractable for DNN-like structures, we instead approximate the probability by sampling from $\theta$ a finite number of times. We can now express the upper equation with a sum over all samples of $\theta$.
\begin{align}
p(y|x,\Theta_{\setminus \theta}) = \sum_{\theta} p(y | x,\Theta) p(\theta)
\end{align}
To sample from $\theta$, we assume that the values are normal distributed with uniform variance and mean centered at the learned weight $w$ and bias $b$:
\begin{align}
\theta &\sim \mathcal{N}(\mu=w,\,\Sigma=I), \mathcal{N}(\mu=b,\,\Sigma=I) \\
\textit{so that }
p(\theta) &= p_{\mathcal{N}(w,I)}(w) \cdot p_{\mathcal{N}(b,I)}(b)
\end{align}

\subsection{Generalizing contributions from classes to tasks}
As proposed by \cite{Robnik-Sikonja2008-of}, we use the weighted evidence ($WE$) to measure the contribution of parameter towards class probability $p(y|x,\Theta)$ (see equation A.6) instead of taking the difference of probabilities directly. $WE_{\theta}(y|x,\Theta)$ gives us a positive value indicating $\theta$ adds evidence for class $y$ for input $x$, while a negative value adds evidence against class $y$ and zero if $\theta$ has no contribution at all. To eventually determine the contribution towards a class independent of the input we calculate the arithmetic mean of the absolute weighted evidence over more than 500 input samples (see equation \ref{eq:class_contribution}) from the test set.
\begin{align}
odds(z) &= \frac{p(z)}{1-p(z)} \\
WE_{\theta}(y|x,\Theta)&=log_2(odds(y|x,\Theta)) \nonumber\\
&- log_2(odds(y|x,\Theta_{\setminus \theta})) \\
C_{\theta}(y|\Theta) &= \frac{1}{n}\sum_{j=1}^n \left| WE_\theta(y|x_j,\Theta) \right| \label{eq:class_contribution}
\end{align}
We finally measure the contribution to a task $t$ by selecting the contributions $C_{\theta}(y|\Theta)$ that satisfy $y = y_{true}$ which are the class predictions that are correct. Furthermore filtering out predictions that had been incorrectly inferred from the network, we can increase certainty that the inputs used to evaluate the contributions lead to high probability for the correct class and low everywhere else. We further generalize the contribution of $\theta$ to task $t$ by averaging over the contributions to each class $y_k$ within task $t$ (see equation \ref{eq:task_contribution}).
\begin{align}
TC_{\theta}(t | \Theta) = \frac{1}{K} \sum_{k=1}^K C_\theta (y_k|\Theta) \label{eq:task_contribution}\\
\nonumber t \in Tasks,\, K \in |outputs_t|
\end{align}

\end{appendix}